\documentclass[12pt]{article}
\usepackage[T1]{fontenc}
\usepackage{lmodern}

\usepackage[letterpaper,margin=1in]{geometry}
\usepackage{amsmath,amssymb,booktabs,graphicx,xcolor}
\usepackage[round,authoryear]{natbib}
\usepackage[hidelinks]{hyperref}
\usepackage{caption}
\newcommand{\Rtail}{R_{\mathrm{tail}}}

\title{Preserving rare-item recall under a storage budget: quantization, distillation, and the limits of teacher fidelity}
\author{Han Chen\textsuperscript{1}\thanks{Corresponding author: \href{mailto:Han.Chen5@utexas.edu}{Han.Chen5@utexas.edu}}
\and Yingrui Li\textsuperscript{2}\\\href{mailto:el516y@gmail.com}{el516y@gmail.com}}
\date{\small\textsuperscript{1}Department of Computer Science, University of Texas at Austin,\\Austin, Texas, United States\\
\textsuperscript{2}Unaffiliated, Fairfax, Virginia, United States}
\begin{document}
\maketitle
\raggedright
\setlength{\parindent}{1.5em}
\begin{abstract}
\raggedright
Recommenders often have to fit their user and item embeddings into a fixed storage budget, for example the memory of an online feature store. The bytes can go to a small full-precision model, a student distilled from a larger teacher, or a higher-rank model stored at lower precision. We asked which choice best preserves the recall of rare (tail) items, and whether a compressed model's fidelity to its teacher shows how well it recalls them. In seven dataset--backbone cells (matrix factorization on Gowalla, Last.fm, Amazon Video Games, Steam, Goodbooks-10k and MovieLens-20M, and LightGCN on Amazon Video Games), we compared seven candidates that fit in 168 bytes per user or item row, the storage of a 40-dimensional full-precision model: a 40-dimensional Bayesian personalized ranking (BPR) model, two 40-dimensional students trained with unbiased knowledge distillation (UnKD), and factorizations of a 300-dimensional teacher's scores at rank 40, 80, 155 and 300, stored in 32, 16, 8 and 4 bits. Each test set was evaluated once, after the candidates and confirmatory analyses had been fixed. The rank-300 4-bit factorization used 13\% of the teacher's storage and kept 95--103\% of its tail recall@20 and 93--97\% of its NDCG@20 in every cell. The same bytes spent on rank 40 gave the lowest tail recall of all candidates in every cell; the 4-bit factorization had 1.2 to 12.7 times its tail recall and 0.99 to 1.46 times its NDCG. The UnKD students kept 56--102\% of the teacher's tail recall, and in all ten prespecified comparisons in the five cells of a prospective expansion, the 4-bit factorization had higher tail recall, with simultaneous 95\% intervals excluding zero. Teacher-hit retention, the share of the teacher's relevant tail hits that a model also makes, ordered 23 of 147 candidate pairs differently from tail recall and identified the wrong tail-recall leader in 3 of 7 cells, because some models find relevant tail items the teacher misses. Spending bytes on rank rather than precision preserved rare-item recall, at 7.5 times the arithmetic per score; compressed recommenders should be judged by total tail recall, not by fidelity to the teacher alone.
\end{abstract}

\section{Introduction}
Most items in a recommender's catalog receive few interactions. Recommending them well matters to users whose tastes popular items do not serve, to the providers of less popular items, and to how much of the catalog people discover~\citep{park,abdollahpouri}. Average accuracy says little about these items. Popular items account for most relevant test interactions, so a model can score well on average while rarely recommending a relevant rare item~\citep{steck}. Rare-item performance therefore has to be measured on its own, and it deserves particular attention when a model is compressed.

User and item embedding tables can account for most of the storage a recommender needs at serving time, and many methods exist to shrink them~\citep{lisurvey,benchmark}. These vectors are often served from an online feature store or an in-memory cache, where storage is paid for every user and item row. Under a fixed storage budget, the bytes can be spent on a small full-precision model, trained directly or distilled from a larger teacher~\citep{rrd,unkd}, or on a larger model stored at lower numeric precision~\citep{guan}. For language models, \citet{dettmers} found that at a fixed number of model bits, 4-bit precision with more parameters generally gives the best accuracy. A similar trade-off is plausible for rare items in recommenders, because low-dimensional embeddings concentrate recommendations on popular items~\citep{ohsaka,lin}. We test it directly, at equal bytes.

A second question is how to judge a compressed model on rare items. A natural yardstick is fidelity to the teacher: the share of the teacher's relevant rare-item recommendations that the compressed model also makes, which we call teacher-hit retention. Fidelity and accuracy need not agree, however~\citep{stanton}. Total rare-item recall, which we call tail recall, also credits relevant items that the compressed model finds and the teacher misses, so the two measures can rank the same models differently. Both require test labels; the question is which of them a compressed model should be judged by.

We compare seven ways of spending one storage budget in seven dataset--backbone combinations, which we call cells: matrix factorization (MF) on Gowalla, Last.fm, Amazon Video Games, Steam, Goodbooks-10k and MovieLens-20M, and LightGCN on Amazon Video Games. The budget is that of a 40-dimensional full-precision model: 168 bytes per user or item row, including an 8-byte ID. The candidates are a 40-dimensional model trained with Bayesian personalized ranking (BPR), two 40-dimensional students trained with unbiased knowledge distillation \citep[UnKD;][]{unkd}, and four low-rank factorizations of a 300-dimensional teacher's scores: rank 40 in 32-bit, rank 80 in 16-bit, rank 155 in 8-bit and rank 300 in 4-bit precision. Two cells come from earlier held-out studies and five from a prospective expansion whose protocol was fixed before any of its data were parsed. In every cell, the candidates and confirmatory analyses were fixed before the test set was evaluated, and the test set was evaluated once.

Three findings stand out (Table~\ref{tab:summary}).
\begin{itemize}
\item \emph{At equal bytes, rank preserved rare-item recall better than precision did.} The rank-300 4-bit factorization, which needs 13.4\% of the teacher's bytes, kept 94.8--102.8\% of the teacher's tail recall@20 and 93.4--97.3\% of its NDCG@20 in every cell. The same bytes spent on a rank-40 FP32 factorization gave the lowest tail recall of the seven candidates in every cell, even where its NDCG stayed close to the teacher's. On MovieLens-20M, for example, it kept 94.6\% of the teacher's NDCG but only 25.3\% of its tail recall.
\item \emph{Distilled students did not match the 4-bit factorization.} The UnKD students kept 56.2--102.4\% of the teacher's tail recall. In the five expansion cells, the 4-bit factorization had higher tail recall than both students in all ten prespecified comparisons, and every simultaneous 95\% interval excluded zero. Only one student, on Gowalla, had confirmed higher tail recall than the 4-bit factorization, and its NDCG was 9.82\% lower.
\item \emph{Teacher fidelity was an unreliable guide to tail recall.} Retention ordered 23 of 147 candidate pairs (15.6\%) differently from tail recall and picked the wrong tail-recall leader in 3 of the 7 cells. On Steam and MovieLens-20M, plain BPR exceeded the teacher's own tail recall, and more than half of its relevant tail hits were items the teacher did not retrieve.
\end{itemize}
Under a storage budget, then, bytes are better spent on rank than on precision when rare items matter, at the price of more arithmetic per score, and compressed recommenders should be judged by the rare items they retrieve, not only by how faithfully they reproduce the teacher.

\subsection{Related work}
\paragraph{Distillation for recommenders.} DE-RRD transfers both a teacher's rankings and its latent knowledge~\citep{rrd}, and DCD corrects discrepancies between the teacher's and the student's rankings~\citep{dcd}. UnKD addresses popularity bias in distillation by sampling teacher-ranked item pairs within popularity groups~\citep{unkd}. RCE-KD analyzes cross-entropy distillation objectives defined over a subset of items~\citep{rce}, and FreqD studies feature-based distillation in terms of graph-spectral frequencies~\citep{freqd}. We study UnKD because it explicitly targets popularity bias, the failure that tail recall exposes.

\paragraph{Compression and quantization.} Benchmarks of lightweight embedding-based recommenders~\citep{benchmark}, surveys of embedding compression~\citep{lisurvey} and product-quantization methods such as RecJPQ and URecJPQ~\citep{recjpq,urecjpq} describe a wide range of alternatives whose effectiveness depends on the dataset. Post-training 4-bit quantization of embedding tables has been applied to production recommendation models~\citep{guan}. Our score factorizations are deliberately simple: they hold everything fixed except rank and precision, so that the two can be compared at equal bytes.

\paragraph{Dimension and popularity.} Low embedding dimension limits what dot-product recommenders can express and biases them toward popular items~\citep{ohsaka}, and item popularity concentrates in the principal spectral component of predicted scores~\citep{lin}. Both analyses predict that truncating a model's rank costs rare items more than popular ones, which is what we observe.

\paragraph{Evaluation practice.} A distilled student's fidelity to its teacher and its generalization can diverge~\citep{stanton}. Offline comparisons of recommenders are sensitive to baselines, tuning and protocol~\citep{dacrema}; sampled-candidate metrics can mislead~\citep{sampled}; and selecting models on the data used to evaluate them biases the results~\citep{cawley}. We therefore rank the full catalog, use explicit tuning and selection rules, and keep model selection separate from held-out evaluation.

\section{Materials and methods}
\subsection{Study design}\label{sec:design}
We compared seven candidates that fit a common storage cap, together with an over-budget teacher, in seven cells (Table~\ref{tab:cells}). Each held-out comparison used five training seeds and ranked the full catalog.

The evidence comes from three study components (Table~\ref{tab:plan}). A development study on Gowalla and MovieLens-1M, which used only validation data, set the training procedure and separated rank from precision with matched-rank controls (Section~\ref{sec:dev}). Earlier held-out studies compared the candidates with MF on Gowalla and Last.fm. A prospective expansion then applied the Last.fm protocol, with its candidates, tuning grids, seeds, margins and statistical analysis, to five new cells: MF on Amazon Video Games, Steam, Goodbooks-10k and MovieLens-20M, and LightGCN on Amazon Video Games. The Last.fm and expansion protocols were fixed before any of their interaction data were parsed. In every cell, the candidates and confirmatory analyses were fixed before the test set was evaluated; later descriptive analyses used the saved outputs.

\begin{table}[tbp]
\centering\footnotesize\setlength{\tabcolsep}{3pt}
\begin{tabular}{lrrrrrrrr}
\toprule
Cell & Users & Items & Train & Valid. & Test & Tail items & Tail test edges & Gini\\
\midrule
Gowalla (MF) & 29,858 & 40,988 & 810,128 & 91,047 & 126,289 & 32,790 & 60,946 & 0.44\\
Last.fm (MF) & 1,859 & 2,823 & 59,367 & 5,719 & 6,269 & 2,258 & 2,107 & 0.60\\
Amazon (MF, LightGCN) & 63,413 & 19,019 & 397,549 & 67,476 & 68,510 & 15,215 & 22,194 & 0.62\\
Steam (MF) & 59,746 & 7,603 & 2,648,039 & 272,262 & 301,530 & 6,082 & 32,953 & 0.85\\
Goodbooks-10k (MF) & 53,366 & 9,999 & 3,383,752 & 349,993 & 388,258 & 7,999 & 123,451 & 0.62\\
MovieLens-20M (MF) & 136,674 & 13,679 & 8,184,945 & 849,642 & 942,859 & 10,943 & 88,551 & 0.86\\
\bottomrule
\end{tabular}
\caption{Evaluated cells and their data. Users and items are those served after filtering; edges are positive user--item interactions in each partition, and test edges are those of eligible users to served items (on Amazon, 12 users had no eligible test edge and are not evaluated). Tail items are the $\lfloor 0.8\,n_i\rfloor$ least frequent items in training; tail test edges are the relevant tail edges $D$ that define tail recall. Gini: Gini coefficient of the items' training frequencies. Each cell's storage cap, the 40-dimensional FP32 payload $(n_u+n_i)(4\cdot 40+8)$ bytes, is given in Tables~\ref{tab:main1} and~\ref{tab:main2}. Gowalla and Last.fm come from the earlier held-out studies; the other cells come from the prospective expansion, and the two Amazon cells share one split.}
\label{tab:cells}
\end{table}

\begin{table}[tbp]
\centering\small\setlength{\tabcolsep}{5pt}
\begin{tabular}{p{2.8cm}p{4.2cm}p{7.1cm}}
\toprule
\raggedright Study component & \raggedright Data and backbones & \raggedright Role in the comparison\tabularnewline
\midrule
\raggedright Development controls & \raggedright Gowalla and MovieLens-1M, MF & \raggedright Validation-only experiments for training settings and matched-rank compression controls.\tabularnewline
\addlinespace[5pt]
\raggedright Earlier held-out studies & \raggedright Gowalla and Last.fm, MF & \raggedright Seven candidates and a teacher, evaluated over five training seeds.\tabularnewline
\addlinespace[5pt]
\raggedright Prospective expansion & \raggedright Amazon Video Games, MF and LightGCN; Steam, Goodbooks-10k and MovieLens-20M, MF & \raggedright Five evaluated cells using the Last.fm candidates, tuning grids, seeds, margins and statistical analysis.\tabularnewline
\bottomrule
\end{tabular}
\caption{Study components and their roles. The seven held-out cells compare the same candidate set under a common storage cap; validation controls separate rank from precision. The Last.fm and expansion protocols were fixed before their interaction data were parsed.}
\label{tab:plan}
\end{table}

\subsection{Datasets and splits}
\paragraph{Gowalla.} We use the Gowalla file of a public dataset collection~\citep{xu2023}, a version of the check-in data~\citep{gowalla} that its publisher had already filtered to a 10-core. We recombine the published partitions and split each user's items at random (seed 2026): 80\% to training, 10\% to validation and the rest to test. The held-out evaluation uses every test user and edge of this split.

\paragraph{Last.fm.} We use the HetRec 2011 Last.fm release~\citep{lastfm}. Every user--artist pair with a positive listening count is a positive; we ignore the counts, remove duplicates and apply an iterative 5-core filter. Each user's interactions are then split by a seeded hash (seed 71007) into training, validation and test edges, approximately 81/9/10, with the same allocation rule as the expansion datasets below.

\paragraph{Expansion datasets.} The four new datasets share one preparation procedure. We keep positive interactions, remove duplicates, and apply a simultaneous iterative 5-core filter to users and items. Each user's $n$ interactions are ordered by a SHA-256 hash of the study, dataset, seed (71007), user and item identifiers; the first $\max(1,\lfloor n/10\rfloor)$ go to test, the next $\max(1,\lfloor (n-n_{\mathrm{test}})/10\rfloor)$ to validation and the rest to training. Only items with at least one training interaction are served, and validation and test edges to other items are dropped (3 validation and 12 test edges on Amazon, 3 and 2 on MovieLens-20M, none elsewhere). The positives are ratings of at least 4 in the raw Video Games reviews of Amazon Reviews 2023~\citep{hou}, with items identified by parent product, which groups product variants; games with positive total playtime in the version-1 Australian user--item data of Steam~\citep{kang,wan,pathak}; ratings of at least 4 in release v1.0 of Goodbooks-10k~\citep{goodbooks}; and ratings of at least 4 in MovieLens-20M~\citep{movielens}. The Amazon MF and LightGCN cells share one split.

\paragraph{MovieLens-1M (validation only).} MovieLens-1M is used only in the development study. We keep ratings of at least 4~\citep{movielens}, remove duplicates, apply a 5-core filter and split each user's interactions 80/20 into training and validation. There is no test split, so the same validation data served both to select models and to report results.

Every filter is applied to the full public interaction graph before splitting, so eligibility is transductive, and no split is temporal.

\subsection{Tail items and metrics}
The tail is the 80\% of served items with the fewest training interactions: we sort the $n_i$ served items by training frequency, breaking ties by item ID, and take the first $\lfloor 0.8\,n_i\rfloor$. At test time, each model ranks the full catalog for every eligible test user after masking that user's training and validation items, and we keep the top 20; ties between FP32 scores are broken by original item ID. No metric uses sampled candidates~\citep{sampled}. A \emph{hit} is a relevant test item in a top-20 list. NDCG@20 is averaged over users. Tail recall and head recall pool hits over all users within a seed: tail recall is the number of tail hits divided by the number of relevant tail test edges. Item-macro tail recall averages per-item recall over the tail items with at least one relevant test edge. Reported means weight the five training seeds (2027--2031) equally. The seeds vary only model training; each dataset has a single split.

\subsection{Retained and off-list hits}\label{sec:decomposition}
For one training seed, let $H_M$ and $H_T$ be the sets of tail user--item hits of a model $M$ and of the teacher. The model's \emph{retained} hits $A_M=H_M\cap H_T$ are those the teacher also makes, and its \emph{off-list} hits $O_M=H_M\setminus H_T$ are relevant tail items outside the teacher's top-20 list. With $D$ the number of relevant tail test edges,
\begin{equation}
 \Rtail(M)=\frac{|A_M|+|O_M|}{D},\qquad
 \mathrm{retention}(M)=\frac{|A_M|}{|H_T|}.
\end{equation}
We call $|A_M|/D$ and $|O_M|/D$ the retained and off-list contributions to tail recall. Within a seed, the retained contribution equals retention times the teacher's own tail recall, $|H_T|/D$. A model can therefore trail another in retention and still lead it in tail recall, if its off-list contribution is large enough.

\subsection{Storage budget and candidates}\label{sec:candidates}
For $n$ combined user and item rows, a $d$-dimensional FP32 payload takes $B_d=n(4d+8)$ bytes: the factors plus one int64 served ID per row, with no bias terms. The common cap in each cell is $B_{40}$, or 168 bytes per row (Tables~\ref{tab:main1} and~\ref{tab:main2}). Integer codecs also store one FP32 scale per row and 8 bytes of codec metadata, so an INT8 rank-155 payload takes $167n+8$ bytes and an INT4 rank-300 payload $162n+8$ bytes. Every candidate fits under the cap without padding, so payloads need not be equal; archive file sizes do not count toward the cap. For a serving store with the same keys, index structure and configuration, fixed per-row overhead is common to the candidates.

The teacher in each cell and seed is a 300-dimensional model of the cell's backbone, stored in FP32 ($1{,}208n$ bytes) and therefore over the cap. Four candidates are factorizations of the teacher's scores. We compute balanced low-rank factors of the teacher's score matrix from thin QR decompositions of its user and item factors, followed by an SVD of the small core matrix, and assign the square roots of the singular values to both sides. These untuned factors are stored at rank 40 in FP32, rank 80 in FP16, rank 155 in INT8 and rank 300 in INT4; we refer to them as FP32-40, FP16-80, INT8 and INT4 (Score-SVD in the result tables). INT4 quantizes the rank-300 factors, not the teacher's original coordinates, and takes 13.4\% of the teacher's bytes. The other three candidates are 40-dimensional FP32 models: a BPR model and two UnKD students. Every codec decodes to FP32 before scoring, so the savings are in stored bytes, not in scoring time or decoded memory. Scoring cost grows with rank: rank 300 needs 7.5 times as many multiply--accumulate operations per scored item as rank 40.

\subsection{Backbones}
MF models are trained with the BPR loss~\citep{bpr}, $-\log\sigma(s_{ui}-s_{uj})$ for a training positive $i$ and a uniformly sampled negative $j$, with Xavier-uniform initialization and an L2 penalty on the sampled embeddings. LightGCN~\citep{lightgcn} uses three propagation layers for teacher and students, averages the embeddings of layers 0 to 3 over the symmetrically normalized training graph, initializes embeddings from $\mathcal{N}(0,0.1^2)$ and uses the softplus form of the same pairwise loss. For LightGCN, the propagated embeddings are both the served factors and the input to the score factorization, and distillation uses the propagated scores. All models are trained in FP32 with Adam and a batch size of 2,048, with TF32 disabled.

\subsection{UnKD students}\label{sec:students}
\paragraph{Gowalla adaptations.} For Gowalla, which was part of the development study, we implemented two students modeled on UnKD~\citep{unkd}, the \emph{source} and \emph{distinct} adaptations. Both use four popularity groups, a sampling temperature of 20, a cache of up to 1,000 teacher-ranked items per group refreshed every epoch, and normal initialization; rounding turns the 30 requested sampled pairs into 32. They differ in one detail of pair sampling: the source adaptation keeps a swap, implemented through a tensor view, that can give both items of a sampled pair the same rank, and the distinct adaptation corrects it. The two were tuned separately, so their selected settings differ as well.

\paragraph{Ports used in the other cells.} For Last.fm and the five expansion cells, we used two ports adapted from the authors' official UnKD code (\url{https://github.com/chengang95/UnKD}, commit 57abd1a). Both keep its initialization; its BPR, distillation and regularization terms; its popularity grouping (two groups of equal popularity mass), caching and sampling rules; and its default of 30 sampled pairs, 15 per group. The \emph{release port} keeps the tensor-view swap of the released code, and the \emph{matched-distinct port} replaces it with a non-aliasing ordering of the same two sampled ranks. Both keep the released sampler's rule of redrawing the whole vector of sampled ranks whenever it contains two equal indices. The distillation loss is added to the BPR and regularization terms with weight $\lambda_{\mathrm{KD}}$. The ports run the released training mechanics on a current software stack and on our data preparation.

\subsection{Training and model selection}
In the Last.fm study and the expansion, each learned method (the teacher, BPR and the two students) had 12 tuning trials with seed 2026. For the teacher and BPR, the grid crosses learning rates $\{10^{-4},10^{-3},10^{-2}\}$ with L2 weights $\{0,10^{-4},10^{-3},10^{-2}\}$; for the students, the same learning rates cross L2 weights $\{10^{-4},10^{-3}\}$ and $\lambda_{\mathrm{KD}}\in\{0.1,1\}$. We selected the converged trial with the highest validation NDCG@20, breaking ties by trial name; tail recall and retention were never used for selection. The teacher's setting was fixed first. For each confirmation seed (2027--2031), a new teacher was trained with that setting, and the students and score factorizations of that seed were derived from it. Each selected setting was used unchanged for all five seeds (Table~\ref{tab:settings}).

Models are evaluated on validation every five epochs. Training stops after 20 evaluations without improvement or at 3,000 epochs, and the checkpoint with the best validation NDCG@20 is kept. Each expansion cell thus required 68 trained models (48 tuning and 20 confirmation fits) and 20 score factorizations. Every confirmation fit converged. Three tuning trials in the Amazon LightGCN cell (one BPR and two release-port trials) reached the 3,000-epoch cap and were therefore ineligible for selection; none had a higher validation NDCG than the selected trial.

The Gowalla development study used different grids: learning rates $\{1,3,10,30\}\times10^{-4}$ and L2 weights $\{0,10^{-5},10^{-3}\}$ for BPR, and learning rates $\{10^{-4},10^{-3}\}$, L2 weights $\{0,10^{-5}\}$ and $\lambda_{\mathrm{KD}}\in\{0.1,1,10\}$ for the two adaptations, with ties broken by tail recall. The Gowalla teacher used a fixed setting carried over from earlier work; all six of its fits converged.

\subsection{Prespecified statistical analysis}
Each cell has its own family of six endpoints. Each of the two UnKD students (the two arms) is compared with INT4 on three quantities: the difference in tail recall, the difference in retention, and the accuracy contrast $A=\mathrm{NDCG}_{\mathrm{arm}}-0.97\,\mathrm{NDCG}_{\mathrm{INT4}}$. Uncertainty comes from 10,000 paired bootstrap resamples of test users (PCG64, seed 91029), shared across methods and across the five fixed seeds. Within each resample, ratio metrics are pooled within a seed and then averaged over seeds with equal weight; seeds are not resampled. Two-sided $p$-values come from the centered bootstrap distribution, counting ties as extreme and applying the $+1$ correction, and are Holm-adjusted across the six endpoints. Simultaneous intervals are basic bootstrap intervals with a Bonferroni correction at a family-wise $\alpha=0.05$. No resample left an endpoint undefined. The intervals describe uncertainty over test users for the fixed trained models; they do not include variation across training seeds, dependence through shared items or dataset shift.

A \emph{reversal}, in which a student beats INT4 on tail recall while trailing it on retention, is confirmed when the interval for the tail-recall difference lies entirely above zero and the interval for the retention difference lies entirely below zero. A practical tail-recall gain requires the lower bound of the tail-recall difference to exceed 0.003 (0.3 percentage points), and noninferiority at the 3\% accuracy margin requires the lower bound of $A$ to be positive. Table~\ref{tab:endpoints} reports all 42 endpoint estimates with their intervals.

All other comparisons are descriptive: the 21 candidate pairs per cell (at the seed mean and for each seed), the per-cell summaries in Table~\ref{tab:summary} and Figures~\ref{fig:ladder} and~\ref{fig:decomp}, item-macro and head recall, and the frequency deciles.

\subsection{Test evaluation and verification}
Test data were kept out of training and model selection. In each cell, a single evaluation ranked the full catalog for all 40 model--seed combinations (seven candidates and the teacher, five seeds each), and no model was refit or reselected afterwards. A separate check verified the output hashes and recomputed the means, intervals and Holm adjustments from the saved statistics. When we optimized the training code, we checked that its sampling and random-number behavior was unchanged.

\subsection{Computing environment}
The expansion ran under Linux on cloud NVIDIA RTX A6000 and A100 (80 GB) GPUs with Python 3.12.13, PyTorch 2.11.0 (CUDA 13.0), NumPy 2.4.4 and SciPy 1.18.0. Some fits moved between the two GPU types and resumed from saved training states. Supplemental Information S1 records further environment details.

\subsection{Use of AI assistance}
OpenAI Codex (GPT-6 Astra and GPT-6.1 Sol) assisted with implementation, optimization, experiment orchestration, data preparation, analysis code, literature checks, drafting of earlier versions, and edits to this version. Anthropic's Claude (configured model claude-opus-5-5, used through the Claude app in October 2026) revised the wording of an earlier version, drafted this version of the manuscript from the verified result tables, computed the derived summaries in Tables~\ref{tab:summary} and~\ref{tab:deciles} and Figures~\ref{fig:ladder} and~\ref{fig:decomp} from those tables, and wrote the figure and table code. The empirical results are metrics computed from the experimental artifacts, not generated observations. Verification relied on file hashes, immutable manifests, checks that resumed runs continued from their saved states, and independent recomputation of the reported statistics from saved outputs. The authors checked the manuscript against the result tables in Supplemental Information S1 and are responsible for all scientific, licensing and formatting claims.

\section{Results}
Tables~\ref{tab:main1} and~\ref{tab:main2} report every candidate and the teacher in every cell, Table~\ref{tab:endpoints} the prespecified endpoints, and Table~\ref{tab:summary} a summary of each cell.

\subsection{Rank versus precision at equal bytes}
\begin{figure}[t]
\centering
\includegraphics[width=\textwidth]{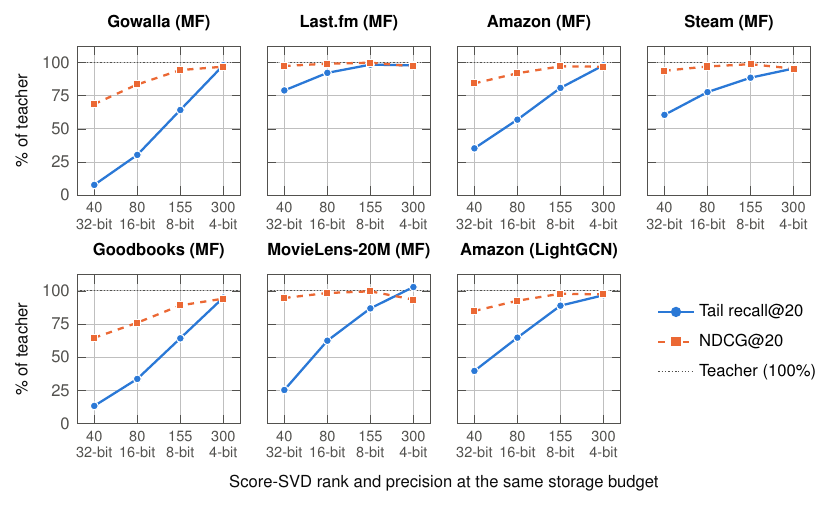}
\caption{Tail recall@20 and NDCG@20 of the four score factorizations as a percentage of the over-budget teacher's, by rank and storage precision. All four factorizations fit the same storage cap. Values are means over five fixed seeds; the dotted line marks the teacher (100\%). Plotted from saved empirical metrics using LaTeX code written with Claude (configured model claude-opus-5-5); the values were checked against Supplemental Information S1.}
\label{fig:ladder}
\end{figure}

Figure~\ref{fig:ladder} follows the four score factorizations, all within the same cap, from rank 40 at 32 bits to rank 300 at 4 bits. Tail recall rose with rank in six of the seven cells; on Last.fm it was nearly flat between rank 155 and rank 300 (98.3\% and 97.7\% of the teacher's). NDCG changed much less. Compared with FP32-40, INT4 had 1.24 (Last.fm) to 12.70 (Gowalla) times the tail recall, but only 0.99 (MovieLens-20M) to 1.46 (Goodbooks-10k) times the NDCG (Table~\ref{tab:summary}). At rank 300 and 4 bits, the factorization kept 94.8--102.8\% of the teacher's tail recall and 93.4--97.3\% of its NDCG.

FP32-40 had the lowest tail recall of all seven candidates in every cell. It uses the same bytes and the same arithmetic per score as the three trained 40-dimensional models, and all three retrieved more tail items. At rank 40, then, training a model directly finds more rare items than truncating a larger one.

At every matched rank in the development study, INT4 NDCG was below FP32 NDCG (Section~\ref{sec:dev}). What 4-bit storage buys under the cap is rank.

\begin{table}[tbp]
\centering\footnotesize\setlength{\tabcolsep}{2.2pt}
\begin{tabular}{llrcrcrccc}
\toprule
Cell & Arm & $\Delta$Tail & 95\% SCI & $\Delta$Ret & 95\% SCI & $A$ & 95\% SCI & Rev. & NI\\
\midrule
Gowalla (MF) & Source & 0.07 & [$-$0.07, 0.20] & $-$11.7 & [$-$12.8, $-$10.6] & $-$1.00 & [$-$1.12, $-$0.88] & No & No\\
 & Distinct & 0.34 & [0.21, 0.47] & $-$9.8 & [$-$11.0, $-$8.8] & $-$0.97 & [$-$1.09, $-$0.85] & Yes & No\\
\addlinespace[2pt]
Last.fm (MF) & Release & $-$0.27 & [$-$0.87, 0.35] & $-$5.9 & [$-$9.6, $-$1.8] & 0.92 & [0.54, 1.33] & No & Yes\\
 & Distinct & $-$0.56 & [$-$1.17, 0.10] & $-$7.4 & [$-$11.2, $-$3.2] & 0.89 & [0.46, 1.32] & No & Yes\\
\addlinespace[2pt]
Amazon (MF) & Release & $-$1.32 & [$-$1.50, $-$1.14] & $-$27.2 & [$-$29.8, $-$24.5] & $-$0.20 & [$-$0.28, $-$0.12] & No & No\\
 & Distinct & $-$1.75 & [$-$1.95, $-$1.55] & $-$33.7 & [$-$36.6, $-$30.8] & $-$0.22 & [$-$0.31, $-$0.14] & No & No\\
\addlinespace[2pt]
Amazon (LightGCN) & Release & $-$0.12 & [$-$0.23, $-$0.01] & $-$8.7 & [$-$10.9, $-$6.4] & 0.18 & [0.12, 0.23] & No & Yes\\
 & Distinct & $-$0.17 & [$-$0.28, $-$0.06] & $-$8.6 & [$-$10.8, $-$6.3] & 0.20 & [0.14, 0.25] & No & Yes\\
\addlinespace[2pt]
Steam (MF) & Release & $-$0.81 & [$-$0.95, $-$0.65] & $-$10.8 & [$-$14.5, $-$7.2] & 1.02 & [0.94, 1.10] & No & Yes\\
 & Distinct & $-$1.02 & [$-$1.17, $-$0.86] & $-$13.5 & [$-$17.3, $-$9.9] & 1.05 & [0.97, 1.12] & No & Yes\\
\addlinespace[2pt]
Goodbooks-10k (MF) & Release & $-$4.11 & [$-$4.25, $-$3.96] & $-$19.7 & [$-$20.3, $-$19.0] & $-$0.99 & [$-$1.07, $-$0.92] & No & No\\
 & Distinct & $-$1.92 & [$-$2.05, $-$1.79] & $-$12.4 & [$-$12.9, $-$11.8] & $-$0.75 & [$-$0.83, $-$0.67] & No & No\\
\addlinespace[2pt]
MovieLens-20M (MF) & Release & $-$0.56 & [$-$0.61, $-$0.50] & $-$21.5 & [$-$23.8, $-$19.3] & 1.56 & [1.51, 1.62] & No & Yes\\
 & Distinct & $-$0.66 & [$-$0.71, $-$0.60] & $-$23.8 & [$-$26.1, $-$21.4] & 1.67 & [1.62, 1.72] & No & Yes\\
\bottomrule
\end{tabular}
\caption{Prespecified endpoints: each UnKD student minus the rank-300 INT4 factorization, in percentage points (100 times the difference in ratios). $\Delta$Tail: tail recall@20; $\Delta$Ret: teacher-hit retention; $A$: NDCG@20 of the student minus 0.97 times that of INT4. SCI: simultaneous 95\% basic bootstrap interval over the cell's six-endpoint family (10,000 paired user resamples, Bonferroni). Rev.: reversal confirmed ($\Delta$Tail interval above zero and $\Delta$Ret interval below zero). NI: noninferiority at the 3\% NDCG margin (lower bound of $A$ above zero). Gowalla uses the development-study source and distinct adaptations; the other cells use the release and matched-distinct ports. Holm-adjusted $p$-values are given in Supplemental Information S1.}
\label{tab:endpoints}
\end{table}

\begin{table}[tbp]
\centering\footnotesize\setlength{\tabcolsep}{3pt}
\begin{tabular}{lrrrrrrrll}
\toprule
 & \multicolumn{2}{c}{INT4 / FP32-40} & \multicolumn{2}{c}{INT4, \% teacher} & \multicolumn{2}{c}{Tail, \% teacher} & & \multicolumn{2}{c}{Leader by}\\
\cmidrule(lr){2-3}\cmidrule(lr){4-5}\cmidrule(lr){6-7}\cmidrule(lr){9-10}
Cell & Tail & NDCG & Tail & NDCG & UnKD & BPR & Mis. & Ret. & Tail\\
\midrule
Gowalla (MF) & 12.70 & 1.41 & 97.8 & 96.7 & 98.7--102.4 & 46.2 & 2 & INT4 & UnKD-D\\
Last.fm (MF) & 1.24 & 1.00 & 97.7 & 97.0 & 92.6--95.3 & 95.6 & 6 & INT8 & INT8\\
Amazon (MF) & 2.78 & 1.15 & 97.8 & 96.7 & 62.7--71.3 & 57.6 & 1 & INT4 & INT4\\
Amazon (LightGCN) & 2.44 & 1.15 & 96.7 & 97.3 & 92.0--93.4 & 69.6 & 4 & INT4 & INT4\\
Steam (MF) & 1.58 & 1.02 & 95.4 & 95.3 & 64.5--71.0 & 108.2 & 6 & INT4 & BPR\\
Goodbooks-10k (MF) & 7.12 & 1.46 & 94.8 & 94.1 & 71.2--83.8 & 66.9 & 1 & INT4 & INT4\\
MovieLens-20M (MF) & 4.07 & 0.99 & 102.8 & 93.4 & 56.2--63.2 & 121.7 & 3 & INT4 & BPR\\
\bottomrule
\end{tabular}
\caption{Per-cell summary of the seven within-budget candidates (means over five fixed seeds). INT4 / FP32-40: ratio of the rank-300 INT4 factorization to the rank-40 FP32 factorization, which fit the same cap. \% of teacher: tail recall@20 (and NDCG@20 for INT4) as a percentage of the over-budget teacher's; the UnKD column spans the two students. Mis.: candidate pairs, of 21, that retention and tail recall order differently; no pair was tied. Leader by: the candidate with the highest retention (Ret.) and the one with the highest tail recall. UnKD-D: distinct student; FP32-40: rank-40 FP32 factorization. All quantities are descriptive.}
\label{tab:summary}
\end{table}

\subsection{Prespecified comparisons: distilled students versus INT4}
In the five expansion cells, INT4 had higher tail recall than both students in all ten prespecified comparisons, by 0.12 to 4.11 percentage points, and every simultaneous interval excluded zero (Table~\ref{tab:endpoints}). The students met the 3\% NDCG noninferiority margin in six of the ten comparisons (Steam, MovieLens-20M and Amazon LightGCN) and missed it in the other four (Amazon MF and Goodbooks-10k).

The two earlier cells were closer. On Last.fm, both ports' tail-recall intervals relative to INT4 included zero, and both ports met the accuracy margin. On Gowalla, the interval for the source adaptation also included zero, whereas the distinct adaptation had 0.34 percentage points higher tail recall than INT4 (interval 0.21 to 0.47). The lower bound of that gain fell short of the prespecified 0.3-point practical margin, and the distinct adaptation's NDCG was 9.82\% lower than that of INT4. Neither Gowalla adaptation met the accuracy margin.

\subsection{Retention as a guide to tail recall}
The prespecified analysis tests directly whether retention can point the wrong way: a confirmed reversal is a student with higher tail recall but lower retention than INT4. One of the fourteen student--INT4 comparisons was a confirmed reversal: the Gowalla distinct adaptation, whose retention was 9.85 percentage points lower than that of INT4 despite its higher tail recall (Table~\ref{tab:endpoints}). None of the five expansion cells had a confirmed reversal: in these cells INT4 led on both measures, with 8.6 to 33.7 points higher retention than the students.

Across all candidates, retention and tail recall agreed on most comparisons but not all (Table~\ref{tab:summary}). Of the 147 candidate pairs (21 per cell), retention ordered 23 (15.6\%) differently from tail recall at the seed mean, with no ties; the count ranged from one pair (Amazon MF and Goodbooks-10k) to six (Last.fm and Steam). Across the 735 seed-level pairs (105 per cell), the two measures disagreed on 111 (15.1\%): 7 on Gowalla, 26 on Last.fm, 3 on Amazon MF, 19 on Amazon LightGCN, 32 on Steam, 5 on Goodbooks-10k and 19 on MovieLens-20M. Seven seed-level pairs, all on Last.fm, were tied on at least one measure.

The disagreements extended to the top candidate. In three cells, the candidate with the highest retention did not have the highest tail recall. On Gowalla, retention favored INT4, whose tail recall was 4.5\% below that of the distinct adaptation. On Steam and MovieLens-20M, retention again favored INT4, but plain BPR had the highest tail recall, and INT4's was 11.9\% and 15.5\% lower. In the other four cells, the retention leader also led on tail recall (INT8 on Last.fm, INT4 elsewhere).

\begin{figure}[t]
\centering
\includegraphics[width=\textwidth]{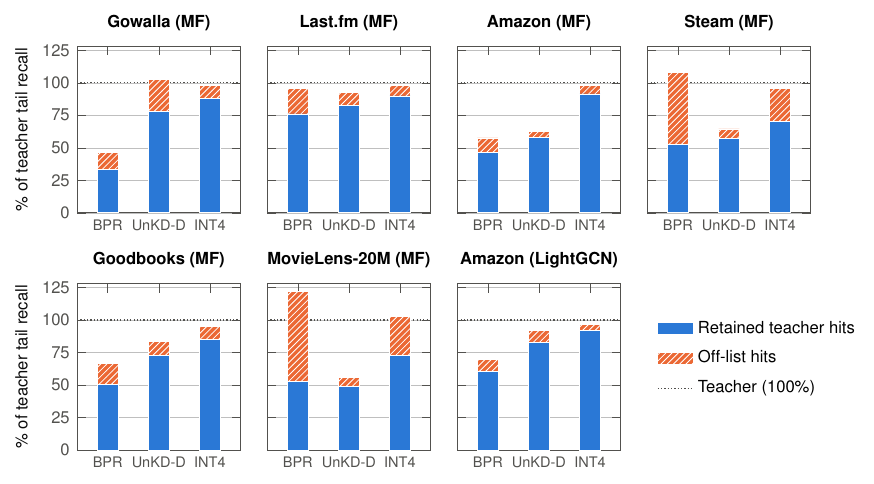}
\caption{Tail recall@20 of BPR, the distinct UnKD student (UnKD-D) and the rank-300 INT4 factorization, split into retained teacher hits (solid) and off-list hits (hatched), as a percentage of the teacher's tail recall. The solid part approximately equals retention. For Gowalla, UnKD-D is the development-study distinct adaptation; elsewhere it is the matched-distinct port. Values are means over five fixed seeds. Plotted from saved empirical metrics using LaTeX code written with Claude (configured model claude-opus-5-5); the values were checked against Supplemental Information S1.}
\label{fig:decomp}
\end{figure}

Figure~\ref{fig:decomp} shows the source of the disagreement. On Steam and MovieLens-20M, BPR retrieved more relevant tail items than the teacher itself (108.2\% and 121.7\% of the teacher's tail recall), and more than half of its tail hits (51.6\% and 56.7\%) were off-list. Retention cannot register these hits. Across the seven cells, off-list hits made up 7.5--23.8\% of the students' tail hits and 5.1--28.9\% of those of INT4.

\subsection{Frequency deciles}
Table~\ref{tab:deciles} breaks recall@20 down by training-frequency decile, from the rarest tenth of items (D1) to the most popular (D10). FP32-40 kept most of the teacher's recall on popular items and little of it on rare ones. In D10 it recalled 85--100\% as much as the teacher in every cell. In the rarest half of the catalog (D1--D5), wherever the teacher's recall was above zero, it recalled at most 27.5\% as much as the teacher in every cell except Last.fm (36--75\%). INT4 recalled 86.4--106.5\% as much as the teacher in every decile where the teacher's recall was at least 1\%. The rarest items were hard for every model: the teacher's recall in D1 was at most 3.0\%, and on Steam and MovieLens-20M the teacher retrieved none of the relevant items in two of the three rarest deciles.

\subsection{Development-study controls}\label{sec:dev}
The development study separated rank from precision with eight ranks (20, 40, 75, 80, 150, 155, 160 and 300), four precisions and five teachers on each of Gowalla and MovieLens-1M. Its 320 validation evaluations were each checked against three storage caps, 20D, 40D and 80D, which 11, 20 and 27 controls per seed fit, respectively. At every matched rank, mean INT4 NDCG was below FP32 NDCG on both datasets. A lower rank can also raise NDCG: on MovieLens-1M, FP32 rank-40 NDCG was 0.302431, against 0.264161 for rank-300 INT4 and 0.281660 for the teacher, consistent with truncation acting as a regularizer~\citep{recjpq}. At the 40D cap, retention and tail recall ordered 24 of 924 strictly comparable seed-level pairs differently on Gowalla and 93 of 929 on MovieLens-1M. On MovieLens-1M, both development adaptations retained more teacher tail hits than BPR yet had lower tail recall in every seed: 0.027025 for the source adaptation and 0.025031 for the distinct adaptation, against 0.030583 for BPR. Because these results reuse the validation data used for selection, they are descriptive.

\subsection{When did distillation help the tail?}\label{sec:exploratory}
Both students had higher NDCG than plain BPR in all seven cells, but higher tail recall only in four: Gowalla, Amazon MF, Amazon LightGCN and Goodbooks-10k. In these cells, BPR reached 46.2--69.6\% of the teacher's tail recall. In the other three (Last.fm, Steam and MovieLens-20M), BPR reached 95.6--121.7\% of it, and both students had lower tail recall than BPR. Distillation may help the tail most when the teacher finds rare items that a small model cannot find on its own. We noticed this pattern after seeing the results, and it rests on seven cells, so we offer it as a hypothesis to test.

\section{Discussion}
\emph{Spend bytes on rank.} Across seven cells, storing a high-rank factorization of the teacher at low precision kept nearly all of the teacher's tail recall. The same bytes spent on rank 40 kept less than half of it in five cells, and in every cell they lost a larger share of tail recall than of NDCG. This parallels the bit-level trade-off reported for language models~\citep{dettmers} and agrees with analyses that link low dimensionality to popularity bias~\citep{ohsaka,lin}. A rank-300 INT4 row requires no more payload bytes to store or transfer than a 40-dimensional FP32 row (162 versus 168 bytes). The trade-off has two costs. Rank 300 needs 7.5 times the arithmetic of rank 40 per scored item once the vectors are decoded, so a system limited by latency rather than storage faces a different choice. And the factorization requires a trained teacher, whose training cost no candidate is charged for. INT4 was also not always the best candidate on the tail: it led in three cells and was overtaken by INT8 on Last.fm, by a student on Gowalla, and by plain BPR on Steam and MovieLens-20M.

\emph{Distillation helped where the teacher had tail recall to give.} Distilling into a 40-dimensional student raised NDCG over plain BPR in every cell, but it raised tail recall only where BPR fell well short of the teacher on the tail. Where BPR already matched or exceeded the teacher there, the students gained few retained hits and lost many of BPR's off-list hits (Tables~\ref{tab:main1} and~\ref{tab:main2}). If the pattern holds more widely, a simple check before distilling for the tail is to compare the small model's validation tail recall with the teacher's.

\emph{Evaluate the tail directly.} Retention asks which of the teacher's relevant recommendations a compressed model preserves. Tail recall asks how many relevant rare items the model retrieves, including those the teacher missed. The difference matters in practice. Retention favored INT4 over a student that retrieved more tail items on Gowalla, and it ranked BPR fifth of seven on Steam and third on MovieLens-20M, the two cells where BPR had the highest tail recall. In each of these cells, the candidate that retention overlooked had more than twice the off-list contribution of INT4. We recommend that evaluations of compressed recommenders on rare items report:
\begin{itemize}
\item the storage cap and the metadata charged against it, the masking rule and the tail denominator;
\item total tail recall and NDCG, with the retained and off-list contributions;
\item rank and precision separately, with matched-rank controls and the complete candidate set;
\item the split, the uncertainty, and the practical and accuracy margins.
\end{itemize}

\subsection{Limitations}\label{sec:limitations}
The evidence is strongest for MF: all six held-out datasets use MF, and only Amazon Video Games also uses LightGCN. Each dataset has a single transductive, non-temporal split, so the results do not address cold-start users or the prediction of future interactions. The tail cutoff (80\% of items) and the list length (20) were fixed in advance and not varied.

We tested one distillation family, UnKD, with 40-dimensional students and 12 tuning trials per method; other distillation methods, student sizes or tuning budgets may behave differently. Because the Gowalla adaptations and the later ports are implemented differently, differences between those cells do not isolate the sampler. On the compression side, we tested score factorization with uniform per-row quantization, not product quantization, hashing or mixed precision, and we did not measure latency.

\section{Conclusions}
Under a fixed storage budget, how the bytes are spent decides how much rare-item recall survives compression. In seven dataset--backbone cells, a rank-300 factorization stored in 4 bits kept 95--103\% of the teacher's tail recall, the same bytes at rank 40 kept the least of any candidate, and no 40-dimensional UnKD student had higher tail recall than the 4-bit factorization in the five expansion cells. Fidelity to the teacher was an unreliable stand-in for tail recall, because some compressed models retrieve relevant rare items that the teacher misses. Reporting total tail recall, split into retained and off-list contributions, alongside NDCG and an explicit storage account makes these differences visible.

\section*{Ethical considerations}
This study reanalyzes public research datasets and reports only aggregate outcomes; we make no attempt to infer user identities or sensitive traits. Rarity here is defined by item popularity, not by any user attribute, so tail recall is not a measure of demographic fairness, and offline relevance does not establish effects on user welfare or item exposure. Access to the datasets remains subject to their providers' terms, including noncommercial-use and attribution conditions. The supplemental material provides aggregate outcomes, saved bootstrap replicates and the Last.fm teacher checkpoint used for validation; the checkpoint contains learned user and item embeddings. It excludes raw and derived interaction arrays, recommendation lists and per-user evaluation statistics; dataset access remains through the original providers.

\section*{Declarations}
\paragraph{Funding.} The authors personally funded this study. No external funding or grant support was received.
\paragraph{Competing interests.} The authors declare no competing interests.
\paragraph{Data availability.} The datasets are publicly available from the sources cited in the manuscript, subject to their providers' terms. Supplemental Information S1 contains the aggregate results for all seven cells: per-seed metrics, inference summaries and the inputs to every table and figure, together with the development-study controls, the fit ledger and the selected settings. It also provides the original 10,000 saved six-endpoint bootstrap replicates for each cell and the observed contrasts, allowing intervals and adjusted p-values to be reconstructed from those saved draws. Per-user inputs needed to regenerate the draws, interaction arrays and recommendation lists are not redistributed.
\paragraph{Code availability.} Supplemental Information S2 contains our code for data preparation, training, compression, evaluation, statistical analysis, tables and figures, with pinned requirements. S1 includes a standalone Python replay of the aggregate tables and code for reconstructing inference from the saved bootstrap replicates. S2 includes the retained Last.fm teacher checkpoint used in the documented CPU validation, which reproduced six validation metrics exactly; raw data must be obtained from the official source, and exact prepared-input requirements are documented in S2. The UnKD students in Last.fm and the expansion cells were trained with code adapted from the authors' official implementation (Chen et al., 2023), available at \url{https://github.com/chengang95/UnKD} (commit 57abd1a327f6e282743c10618511f21718ef0232). Because that repository has no license, S2 does not include the adapted file; UNKD\_REBUILD.md documents our changes relative to the pinned commit.
\paragraph{Author contributions.} Han Chen: Conceptualization, Methodology, Software, Formal analysis, Investigation, Data curation, Resources, and Writing -- original draft. Yingrui Li: Conceptualization, Methodology, Validation, Formal analysis, Investigation, Data curation, Resources, Visualization, and Writing -- review \& editing. Both authors approved the final manuscript.
\paragraph{Human participants and animals.} The study reanalyzes previously released public research datasets. We recruited no participants, collected no new personal data and conducted no animal research.
\paragraph{Supplemental information.} Supplemental Information S1: aggregate results for all seven cells, the inputs to every table and figure, development-study controls, fit and selection records, environment records, aggregate-table replay, and saved bootstrap replicates with inference-reconstruction code. Supplemental Information S2: first-party experimental and reporting code, pinned requirements, validation checks, the retained Last.fm validation checkpoint and instructions for rebuilding the UnKD ports.

\bibliographystyle{plainnat}
\bibliography{references}

\clearpage
\begin{table}[p]
\centering\footnotesize\setlength{\tabcolsep}{3.5pt}
\begin{tabular}{lrrrrrr}
\toprule
Method & Payload bytes & NDCG@20 & Tail R@20 & Ret.\ (\%) & Retained / $D$ & Off-list / $D$\\
\midrule
\multicolumn{7}{l}{\emph{Gowalla (MF); common cap 11,902,128 bytes}}\\
BPR 40D & 11,902,128 & 11.01 & 3.42 & 33.1 & 2.45 & 0.97\\
UnKD source adapt.\ 40D & 11,902,128 & 12.82 & 7.32 & 76.2 & 5.65 & 1.67\\
UnKD distinct adapt.\ 40D & 11,902,128 & 12.85 & 7.59 & 78.0 & 5.78 & 1.81\\
Score-SVD 40D FP32 & 11,902,128 & 10.10 & 0.57 & 5.9 & 0.44 & 0.13\\
Score-SVD 80D FP16 & 11,902,128 & 12.27 & 2.24 & 26.5 & 1.96 & 0.28\\
Score-SVD 155D INT8 & 11,831,290 & 13.87 & 4.75 & 59.5 & 4.41 & 0.34\\
Score-SVD 300D INT4 & 11,477,060 & 14.24 & 7.25 & 87.9 & 6.51 & 0.74\\
Teacher 300D FP32 (over cap) & 85,581,968 & 14.73 & 7.41 & 100.0 & 7.41 & 0.00\\
\midrule
\multicolumn{7}{l}{\emph{Last.fm (MF); common cap 786,576 bytes}}\\
BPR 40D & 786,576 & 26.53 & 10.42 & 76.0 & 8.30 & 2.13\\
UnKD release 40D & 786,576 & 26.94 & 10.39 & 84.0 & 9.17 & 1.22\\
UnKD matched-distinct 40D & 786,576 & 26.91 & 10.10 & 82.5 & 9.00 & 1.10\\
Score-SVD 40D FP32 & 786,576 & 26.90 & 8.60 & 75.1 & 8.19 & 0.41\\
Score-SVD 80D FP16 & 786,576 & 27.38 & 10.04 & 90.2 & 9.83 & 0.21\\
Score-SVD 155D INT8 & 781,902 & 27.58 & 10.72 & 97.5 & 10.64 & 0.08\\
Score-SVD 300D INT4 & 758,492 & 26.83 & 10.66 & 89.9 & 9.81 & 0.85\\
Teacher 300D FP32 (over cap) & 5,655,856 & 27.66 & 10.91 & 100.0 & 10.91 & 0.00\\
\midrule
\multicolumn{7}{l}{\emph{Amazon (MF); common cap 13,848,576 bytes}}\\
BPR 40D & 13,848,576 & 6.29 & 2.87 & 46.3 & 2.31 & 0.56\\
UnKD release 40D & 13,848,576 & 7.34 & 3.56 & 64.2 & 3.20 & 0.35\\
UnKD matched-distinct 40D & 13,848,576 & 7.32 & 3.13 & 57.7 & 2.88 & 0.25\\
Score-SVD 40D FP32 & 13,848,576 & 6.77 & 1.76 & 30.0 & 1.50 & 0.26\\
Score-SVD 80D FP16 & 13,848,576 & 7.38 & 2.84 & 51.4 & 2.56 & 0.27\\
Score-SVD 155D INT8 & 13,766,152 & 7.80 & 4.03 & 76.5 & 3.82 & 0.21\\
Score-SVD 300D INT4 & 13,353,992 & 7.77 & 4.88 & 91.4 & 4.56 & 0.32\\
Teacher 300D FP32 (over cap) & 99,577,856 & 8.04 & 4.99 & 100.0 & 4.99 & 0.00\\
\midrule
\multicolumn{7}{l}{\emph{Amazon (LightGCN); common cap 13,848,576 bytes}}\\
BPR 40D & 13,848,576 & 6.51 & 2.52 & 60.4 & 2.18 & 0.33\\
UnKD release 40D & 13,848,576 & 7.66 & 3.38 & 83.0 & 3.00 & 0.38\\
UnKD matched-distinct 40D & 13,848,576 & 7.68 & 3.33 & 83.1 & 3.01 & 0.32\\
Score-SVD 40D FP32 & 13,848,576 & 6.72 & 1.43 & 36.8 & 1.33 & 0.10\\
Score-SVD 80D FP16 & 13,848,576 & 7.34 & 2.34 & 62.0 & 2.24 & 0.10\\
Score-SVD 155D INT8 & 13,766,152 & 7.75 & 3.21 & 87.2 & 3.15 & 0.06\\
Score-SVD 300D INT4 & 13,353,992 & 7.71 & 3.50 & 91.7 & 3.32 & 0.18\\
Teacher 300D FP32 (over cap) & 99,577,856 & 7.93 & 3.62 & 100.0 & 3.62 & 0.00\\
\bottomrule
\end{tabular}
\caption{Held-out results for all candidates and the over-budget teacher (means over five fixed seeds). NDCG, tail recall and the retained and off-list contributions are multiplied by 100; retention is a percentage. The retained and off-list contributions sum to tail recall before rounding. Payload bytes count embeddings, served row IDs and codec metadata. On Gowalla the two students are the development-study source and distinct adaptations; elsewhere they are the release and matched-distinct ports. Per-seed values, standard deviations and secondary metrics are given in Supplemental Information S1.}
\label{tab:main1}
\end{table}

\begin{table}[p]
\centering\footnotesize\setlength{\tabcolsep}{3.5pt}
\begin{tabular}{lrrrrrr}
\toprule
Method & Payload bytes & NDCG@20 & Tail R@20 & Ret.\ (\%) & Retained / $D$ & Off-list / $D$\\
\midrule
\multicolumn{7}{l}{\emph{Steam (MF); common cap 11,314,632 bytes}}\\
BPR 40D & 11,314,632 & 26.30 & 3.58 & 52.4 & 1.73 & 1.84\\
UnKD release 40D & 11,314,632 & 28.10 & 2.34 & 59.8 & 1.97 & 0.37\\
UnKD matched-distinct 40D & 11,314,632 & 28.13 & 2.13 & 57.1 & 1.89 & 0.25\\
Score-SVD 40D FP32 & 11,314,632 & 27.46 & 2.00 & 33.5 & 1.11 & 0.89\\
Score-SVD 80D FP16 & 11,314,632 & 28.40 & 2.56 & 49.2 & 1.63 & 0.94\\
Score-SVD 155D INT8 & 11,247,291 & 28.91 & 2.92 & 66.9 & 2.21 & 0.71\\
Score-SVD 300D INT4 & 10,910,546 & 27.93 & 3.15 & 70.6 & 2.33 & 0.82\\
Teacher 300D FP32 (over cap) & 81,357,592 & 29.31 & 3.30 & 100.0 & 3.30 & 0.00\\
\midrule
\multicolumn{7}{l}{\emph{Goodbooks-10k (MF); common cap 10,645,320 bytes}}\\
BPR 40D & 10,645,320 & 17.94 & 11.62 & 50.6 & 8.79 & 2.83\\
UnKD release 40D & 10,645,320 & 23.10 & 12.35 & 65.8 & 11.42 & 0.93\\
UnKD matched-distinct 40D & 10,645,320 & 23.35 & 14.54 & 73.1 & 12.69 & 1.85\\
Score-SVD 40D FP32 & 10,645,320 & 17.06 & 2.31 & 11.6 & 2.02 & 0.30\\
Score-SVD 80D FP16 & 10,645,320 & 20.06 & 5.84 & 30.4 & 5.28 & 0.56\\
Score-SVD 155D INT8 & 10,581,963 & 23.50 & 11.15 & 60.3 & 10.46 & 0.68\\
Score-SVD 300D INT4 & 10,265,138 & 24.84 & 16.46 & 85.5 & 14.83 & 1.63\\
Teacher 300D FP32 (over cap) & 76,544,920 & 26.41 & 17.36 & 100.0 & 17.36 & 0.00\\
\midrule
\multicolumn{7}{l}{\emph{MovieLens-20M (MF); common cap 25,259,304 bytes}}\\
BPR 40D & 25,259,304 & 22.38 & 1.72 & 52.7 & 0.74 & 0.97\\
UnKD release 40D & 25,259,304 & 23.55 & 0.89 & 51.6 & 0.73 & 0.16\\
UnKD matched-distinct 40D & 25,259,304 & 23.66 & 0.79 & 49.3 & 0.70 & 0.10\\
Score-SVD 40D FP32 & 25,259,304 & 22.96 & 0.36 & 17.8 & 0.25 & 0.11\\
Score-SVD 80D FP16 & 25,259,304 & 23.87 & 0.88 & 43.3 & 0.61 & 0.27\\
Score-SVD 155D INT8 & 25,108,959 & 24.19 & 1.22 & 65.9 & 0.93 & 0.30\\
Score-SVD 300D INT4 & 24,357,194 & 22.67 & 1.45 & 73.1 & 1.03 & 0.42\\
Teacher 300D FP32 (over cap) & 181,626,424 & 24.28 & 1.41 & 100.0 & 1.41 & 0.00\\
\bottomrule
\end{tabular}
\caption{Held-out results, continued (same layout as Table~\ref{tab:main1}).}
\label{tab:main2}
\end{table}

\begin{table}[p]
\centering\footnotesize\setlength{\tabcolsep}{4pt}
\begin{tabular}{lrrrrrrrrrr}
\toprule
 & D1 & D2 & D3 & D4 & D5 & D6 & D7 & D8 & D9 & D10\\
\midrule
\multicolumn{11}{l}{\emph{Gowalla (MF)}}\\
Teacher & 2.11 & 3.89 & 4.93 & 5.80 & 7.32 & 7.90 & 10.67 & 13.39 & 17.27 & 33.19\\
FP32-40 & 0.01 & 0.06 & 0.07 & 0.16 & 0.27 & 0.43 & 0.82 & 1.90 & 3.96 & 30.71\\
INT4 & 2.13 & 3.85 & 4.86 & 5.76 & 7.22 & 7.84 & 10.38 & 12.89 & 16.80 & 32.14\\
BPR & 0.40 & 1.13 & 1.51 & 1.88 & 2.96 & 3.59 & 5.36 & 7.84 & 11.37 & 28.38\\
UnKD-D & 2.23 & 4.42 & 6.18 & 7.47 & 9.47 & 7.64 & 9.27 & 12.49 & 13.93 & 29.48\\
\addlinespace[3pt]
\multicolumn{11}{l}{\emph{Last.fm (MF)}}\\
Teacher & 2.98 & 5.00 & 5.67 & 9.26 & 11.40 & 10.24 & 14.27 & 16.25 & 26.36 & 61.03\\
FP32-40 & 1.08 & 2.37 & 3.00 & 6.29 & 8.50 & 7.24 & 12.44 & 14.34 & 25.48 & 61.30\\
INT4 & 2.58 & 4.47 & 5.33 & 9.49 & 10.90 & 9.61 & 14.56 & 15.96 & 25.46 & 59.94\\
BPR & 1.63 & 3.16 & 4.50 & 8.69 & 11.30 & 9.45 & 15.36 & 15.79 & 26.21 & 58.64\\
UnKD-D & 1.29 & 3.16 & 4.00 & 7.66 & 8.80 & 8.43 & 13.64 & 17.88 & 35.85 & 57.30\\
\addlinespace[3pt]
\multicolumn{11}{l}{\emph{Amazon (MF)}}\\
Teacher & 0.64 & 1.63 & 2.48 & 2.56 & 3.24 & 4.67 & 6.95 & 9.06 & 10.51 & 26.00\\
FP32-40 & 0.02 & 0.07 & 0.08 & 0.32 & 0.30 & 1.17 & 2.64 & 4.36 & 5.41 & 25.59\\
INT4 & 0.68 & 1.55 & 2.50 & 2.41 & 3.25 & 4.60 & 6.73 & 8.86 & 10.25 & 25.23\\
BPR & 0.17 & 0.37 & 0.67 & 0.54 & 1.36 & 2.43 & 4.32 & 6.15 & 7.73 & 22.91\\
UnKD-D & 0.23 & 0.60 & 0.91 & 0.94 & 1.31 & 2.45 & 4.51 & 6.76 & 9.20 & 24.67\\
\addlinespace[3pt]
\multicolumn{11}{l}{\emph{Amazon (LightGCN)}}\\
Teacher & 0.35 & 0.63 & 1.20 & 1.15 & 1.96 & 2.77 & 5.43 & 7.42 & 8.70 & 27.47\\
FP32-40 & 0.02 & 0.00 & 0.01 & 0.09 & 0.39 & 0.74 & 1.90 & 3.87 & 5.16 & 25.75\\
INT4 & 0.35 & 0.65 & 1.16 & 1.18 & 1.86 & 2.67 & 5.24 & 7.16 & 8.40 & 26.88\\
BPR & 0.21 & 0.36 & 0.50 & 0.54 & 1.18 & 1.68 & 3.84 & 5.56 & 6.48 & 23.81\\
UnKD-D & 0.27 & 0.60 & 0.88 & 0.94 & 1.54 & 2.30 & 4.79 & 7.33 & 9.99 & 26.55\\
\addlinespace[3pt]
\multicolumn{11}{l}{\emph{Steam (MF)}}\\
Teacher & 0.00 & 0.00 & 0.56 & 0.95 & 1.46 & 2.93 & 3.79 & 4.39 & 5.48 & 39.28\\
FP32-40 & 0.00 & 0.00 & 0.00 & 0.05 & 0.40 & 1.09 & 2.46 & 2.97 & 3.08 & 37.96\\
INT4 & 0.00 & 0.07 & 1.06 & 1.23 & 1.55 & 2.79 & 3.62 & 4.06 & 5.17 & 37.68\\
BPR & 0.00 & 0.00 & 0.16 & 0.55 & 0.98 & 2.44 & 3.88 & 5.31 & 6.56 & 35.32\\
UnKD-D & 0.00 & 0.02 & 0.23 & 0.59 & 0.80 & 1.75 & 2.42 & 2.93 & 4.29 & 37.80\\
\addlinespace[3pt]
\multicolumn{11}{l}{\emph{Goodbooks-10k (MF)}}\\
Teacher & 1.80 & 4.94 & 6.46 & 9.99 & 14.67 & 19.97 & 23.36 & 24.51 & 29.60 & 39.64\\
FP32-40 & 0.11 & 0.50 & 0.51 & 0.53 & 0.89 & 1.77 & 2.71 & 4.95 & 9.81 & 33.52\\
INT4 & 1.89 & 4.64 & 6.26 & 9.52 & 13.74 & 18.85 & 22.15 & 23.30 & 28.41 & 37.58\\
BPR & 3.02 & 4.48 & 5.56 & 6.99 & 9.87 & 13.35 & 14.85 & 15.87 & 19.85 & 28.09\\
UnKD-D & 1.74 & 3.93 & 5.00 & 7.72 & 11.22 & 16.44 & 19.63 & 21.47 & 27.57 & 34.15\\
\addlinespace[3pt]
\multicolumn{11}{l}{\emph{MovieLens-20M (MF)}}\\
Teacher & 0.00 & 0.07 & 0.00 & 0.03 & 0.20 & 0.48 & 0.99 & 2.28 & 5.07 & 31.59\\
FP32-40 & 0.00 & 0.00 & 0.00 & 0.00 & 0.00 & 0.02 & 0.14 & 0.66 & 2.51 & 30.57\\
INT4 & 0.00 & 0.00 & 0.00 & 0.02 & 0.12 & 0.52 & 1.07 & 2.33 & 4.98 & 29.69\\
BPR & 0.00 & 0.00 & 0.00 & 0.02 & 0.13 & 0.41 & 1.24 & 2.83 & 5.72 & 29.33\\
UnKD-D & 0.00 & 0.00 & 0.09 & 0.05 & 0.21 & 0.28 & 0.51 & 1.28 & 3.78 & 30.77\\
\bottomrule
\end{tabular}
\caption{Recall@20 ($\times 100$) by training-frequency decile, from the rarest tenth of items (D1) to the most popular (D10); means over five fixed seeds. Deciles follow training frequency with ties broken by item ID, so the 80\% tail cutoff need not align with a decile boundary. FP32-40: rank-40 FP32 factorization; INT4: rank-300 INT4 factorization; UnKD-D: distinct student (on Gowalla, the development-study distinct adaptation). The other candidates' deciles are given in Supplemental Information S1.}
\label{tab:deciles}
\end{table}

\begin{table}[p]
\centering\footnotesize\setlength{\tabcolsep}{5pt}
\begin{tabular}{lllll}
\toprule
Cell & Teacher & BPR & UnKD release (source) & UnKD distinct\\
\midrule
Gowalla (MF) & $10^{-4}$, $10^{-4}$ & $3{\times}10^{-4}$, $10^{-3}$ & $10^{-4}$, 0, 1 & $10^{-4}$, $10^{-5}$, 0.1\\
Last.fm (MF) & $10^{-3}$, $10^{-2}$ & $10^{-3}$, $10^{-2}$ & $10^{-4}$, $10^{-4}$, 1 & $10^{-4}$, $10^{-4}$, 1\\
Amazon (MF) & $10^{-4}$, $10^{-3}$ & $10^{-3}$, $10^{-2}$ & $10^{-4}$, $10^{-3}$, 1 & $10^{-4}$, $10^{-3}$, 1\\
Amazon (LightGCN) & $10^{-3}$, $10^{-3}$ & $10^{-3}$, $10^{-3}$ & $10^{-4}$, $10^{-4}$, 1 & $10^{-4}$, $10^{-4}$, 1\\
Steam (MF) & $10^{-4}$, $10^{-3}$ & $10^{-4}$, $10^{-3}$ & $10^{-4}$, $10^{-4}$, 1 & $10^{-4}$, $10^{-4}$, 1\\
Goodbooks-10k (MF) & $10^{-4}$, $10^{-3}$ & $10^{-4}$, 0 & $10^{-4}$, $10^{-3}$, 1 & $10^{-4}$, $10^{-4}$, 0.1\\
MovieLens-20M (MF) & $10^{-4}$, $10^{-3}$ & $10^{-4}$, $10^{-3}$ & $10^{-4}$, $10^{-4}$, 1 & $10^{-4}$, $10^{-4}$, 1\\
\bottomrule
\end{tabular}
\caption{Selected settings: learning rate and L2 weight, plus $\lambda_{\mathrm{KD}}$ for the students. Each was chosen on seed 2026 by validation NDCG@20 and used unchanged for the five confirmation seeds; on Gowalla the student columns are the source and distinct adaptations, and the teacher's setting was carried over from an earlier study. Grids and every trial's outcome are in the fit ledger in Supplemental Information S1.}
\label{tab:settings}
\end{table}

\end{document}